# Elastic anisotropy and Surface Acoustic Wave propagation in CoFeB/Au multilayers: influence of thickness and light penetration depth

A. V. Achuthan[1], S. Janardhanan[1], P. Kuświk[2], A. Trzaskowska[1*]

[1] *ISQI, Faculty of Physics and Astronomy, Adam Mickiewicz University,*

*Uniwersytetu Poznańskiego 2, 61-614 Poznan, Poland*

[2] *Institute of Molecular Physics, Polish Academy of Science, Smoluchowskiego 17,*

*60-179 Poznan, Poland.*

* E-mail: olatrzas@amu.edu.pl

**Abstract**

Surface acoustic waves (SAWs) in multilayered nanostructures represent a critical frontier in understanding material behavior at the nanoscale, with profound implications for emerging acoustic and spintronic technologies. In this study, we investigate the influence of the magnetic layer thickness on the propagation of surface acoustic waves in CoFeB-based multilayers. Two approaches to effective medium modelling are considered: one treating the entire multilayer as a homogeneous medium and another focusing on the region affected by light penetration. The elastic properties of the system are analyzed using Brillouin light scattering and numerical modelling, with a particular emphasis on the anisotropy of Young's modulus and its dependence on CoFeB thickness. The results reveal a significant variation in surface acoustic wave velocity and elastic anisotropy as a function of the multilayer configuration, highlighting the role of the penetration depth in effective medium approximations. These findings provide valuable insights into the tunability of acoustic and spin-wave frequencies through structural modifications, which is crucial for the development of high-performance resonators, surface acoustic wave filters, and spin-wave-based information processing devices.

## 1. Introduction

Surface acoustic waves (SAW) have intrigued researchers for centuries, with significant contributions to the field dating back to 1887, when Lord Rayleigh initiated studies on this phenomenon. SAWs are elastic waves that propagate along the surface of a material, with their intensity exponentially decaying with depth. There are different types of surface acoustic waves, including Rayleigh, Sezawa, Lamb, and Love waves. A well-known mode is the Rayleigh surface acoustic wave (R-SAW), which can propagate along the free surface of a semi-infinite solid. In homogenous materials, R-SAWs are the most predominant waves that exist there. However, for nonhomogeneous materials for example substrate with the layer on them different types of surface wave can propagate[1]. Rayleigh surface waves exhibit distinct behaviors in isotropic versus anisotropic materials. In isotropic materials, R-SAWs are dispersive[2], while in anisotropic materials, the amplitude decays with depth in an oscillatory manner. Importantly, the properties of Rayleigh waves, including phase velocity, are highly dependent on the direction of propagation relative to the crystallographic orientation of the studied material. Additionally, the presence of layers can significantly alter the propagation characteristics of R-SAW, adding complexity to their behavior in anisotropic materials[1,3]. The dispersion relations of SAWs can be substantially modified through various means, including the use of phononic crystals, spatial constraints, or external stress fields[4].

Determining the velocity, dispersion, and anisotropy of SAWs offers a non-invasive, highly sensitive method for accurately measuring elastic parameters, allowing researchers to determine nonlinear elastic properties at interfaces between rough surfaces and assess the elastic characteristics of thin films. Moreover, studies on SAWs in magnetic thin films have provided valuable insights into magnon-phonon interactions, an essential element for advancing multifunctional spintronic devices. The implications of SAW research are extensive, with significant potential to drive innovation in materials science, spintronics, and nanotechnology, emphasizing the importance of ongoing research in this area[5,6]. Through numerous experimental investigations, SAWs have demonstrated extensive applicability across various material systems. Their unique characteristics, such as high resistivity[7], low power consumption[8], and compatibility with integrated circuit technology[9] have facilitated their adoption in numerous applications. Notable examples include surface flaw detection[10], ultrasonic signal processing devices, Fourier-transform processors[11], SAW sensors, especially for biological and chemical/microfluidic-based sensing applications[12,13] structural health monitoring, telecommunications, damage detection in metallic structures[14], wireless passive thermometer[15], communication applications[16] and so on.

In the current technology for microsystem fabrication, silicon is the most employed material, on which various layers with distinct properties are deposited. From an application standpoint, a thorough understanding of the fundamental properties of such systems is therefore crucial. One such property is the elastic characteristics, which can be indirectly determined through investigations into SAW propagation. The importance of surface waves in modern electronics, spintronics, and phonon engineering highlights the role of multilayer systems, particularly magnetic structures, in SAW propagation and spin wave phenomena. Understanding the fundamental properties of multilayer structures, including elastic characteristics, is essential for the continued advancement of technology.

One of the methods used to determine SAW properties, such as dispersion and velocity anisotropy, is Brillouin light scattering (BLS)[17]. Due to its high-frequency resolution, flexibility with samples, and localized spatial capabilities, BLS emerges as a powerful technique for measuring material properties[18]. Notably, BLS is the only technique capable of studying SAW dynamics, as well as the propagation of longitudinal and transverse bulk waves and pseudo-surface waves in the GHz frequency range[4]. It is a well-established technique frequently employed in nondestructive testing to assess the elastic properties of bulk materials and thin films[19].

In this study, we investigated thermally excited surface acoustic waves in Si/Ti/Au/CoFeB/Au heterostructures using BLS techniques, which are of particular interest for applications in memory media and magnetoresistive sensors. Previous studies have demonstrated that CoFeB/Au layers exhibit Dzyaloshinskii-Moriya interaction (DMI), and their magnetic properties can be significantly modified by adjusting the thickness of the layers [5]. This leads to a pertinent question regarding the behavior of surface acoustic waves in response to variations in the thickness of the CoFeB layer. This study employs BLS techniques to investigate thermally excited surface acoustic waves in Si/Ti/Au/CoFeB/Au heterostructures, which hold promise for applications in memory media and magnetoresistive sensors. Given the impact of the DMI on the magnetic properties of CoFeB/Au layers and their sensitivity to layer thickness, we explore how variations in CoFeB thickness affect surface acoustic wave behavior.

Despite the intriguing magnetic properties of CoFeB layers, the phononic characteristics of these multilayer systems have received insufficient attention. To address this gap, we employed high-resolution Brillouin light scattering spectroscopy to investigate the propagation of Rayleigh and Sezawa waves within these heterostructures. Our experimental findings were validated and complemented by numerical simulations using COMSOL Multiphysics software, which enabled us to estimate key elastic parameters such as Young's modulus and Zener anisotropy parameters. This integrated approach provides a deeper understanding of the relationships between magnetic and acoustic properties in these complex heterostructures.

## 2. Materials and methods

### 2.1. The crystal

The Ti(4 nm)/Au(60 nm)/CoFeB($t_{CoFeB}$)/Au(2 nm) sample with different thicknesses of CoFeB ($t_{CoFeB}$ = 0.8 – 2 nm) was deposited onto naturally oxidized Si (001) substrate using magnetron sputtering in Ar atmosphere at $P_{Ar}$ = 1.4 × $10^{-3}$ mbar. The deposition was performed with base pressure < 2 × $10^{-8}$ mbar. The dimensions of the sample were 5 × 10 mm$^2$. The CoFeB layer was sputtered from a $Co_{20}Fe_{60}B_{20}$ target, the composition of which was earlier verified by energy dispersive X-ray spectroscopy. The thicknesses of the Ti, Au, and CoFeB layers were controlled by selecting the appropriate deposition time, based on the deposition rate obtained from profilometer measurements for the calibration sample. The thickness of the individual components forming the multilayer material has been placed in parentheses next to each material. An amorphous phase of CoFeB was verified by an x-ray diffractometer in grazing incident configuration[20].

### 2.2. Experimental setup

Using BLS technique we can measures inelastic scattering between incident photons and thermal phonons (acoustic waves). Phonon characteristics are determined by measuring the incident light's wavevector projection ($q$) and scattered light's frequency shift ($\Delta f$). Momentum conservation dictates that the acoustic phonon's wavevector is equivalent to the in-plane projection of the incident light's wavevector.
In our study, surface acoustic phonon propagation was investigated using a six-pass tandem Brillouin spectrometer (JRS Scientific Instruments), which provides a contrast of $10^{15}$ [21]. The source of light was a 3.5W laser Coherent V6 emitting the second harmonics of light of the length $\lambda_0$ = 532 nm. Brillouin Light Scattering experiments measure the relative frequency shifts, represented by Stokes and anti-Stokes components, that occur when laser light undergoes inelastic scattering by acoustic phonons. A detailed description of the experimental setup is found in Refs.[20,21]. The measurements were conducted in the backscattering geometry with *pp* polarization for both incident and scattered light. Both polarizations were confined to the sagittal plane of the sample, defined by the wave vector of the phonon and the normal to the sample surface.
The wavevector $q$ is given by the following equation:

$$q = \frac{4\pi \sin\theta}{\lambda_0} \qquad (1)$$

where $\theta$ is the angle between the incident light and the normal of the sample, $\lambda_0$ is the wavelength of the laser light. As $\theta$ varies from $5^0$ to $85^0$ value of wavevector varies from 2 to 22.7 μm$^{-1}$.

The correlation between $\theta$ angle and frequency of the SAW gives us phase velocity $\upsilon$ which can be found in the equation:

$$\upsilon = \frac{\Delta f \lambda_0}{2\sin\theta} \qquad (2)$$

Brillouin spectroscopy's high sensitivity enables detailed characterization of the dispersion relation governing SAW propagation in the samples under study.

## 3. Finite-element method simulations

To investigate surface acoustic waves, we employed COMSOL Multiphysics software[22], which utilizes the finite-element method (FEM) to tackle complex coupled systems of partial differential equations. We used frequency-domain study to reflect the experimental excitation of the system. In the materials analyzed, the intrinsic inhomogeneity, variable density, and specific attributes of the elastic tensor play a crucial role in determining the localization of surface modes and, consequently, the dispersion relations of the systems.

For a precise assessment of mode localization and dispersion characteristics, we modelled the silicon substrate as a uniform semi-infinite medium ($z \leq 0$), with nanostructures positioned on its surface. To maintain consistency in our analysis, we assumed the substrate to be perfectly flat during the simulations.

The elastic tensor and mass density for the CoFeB were calculated using the experimental values of the elastic tensor for each component (cobalt[23–25], iron[26,27] and boron[28,29]) at room temperature with respect to the chemical concentration.

The calculations were performed for the elastic constants of all components which create the sample - look at table 1.

Table 1. Elastic properties $c_{ij}$ (GPa) and density $\rho$ (kg/m$^3$) of materials used in FEM simulations.

|        | Silicon [30] | Titanium [31] | Gold [30] | CoFeB [31] |
|--------|--------------|---------------|-----------|------------|
| $c_{11}$ | 165.7        | 178           | 190       | 267        |
| $c_{12}$ | 63.9         | 80.9          | 161       | 85         |
| $c_{44}$ | 79.9         | 43.4          | 42.3      | 120        |
| $\rho$   | 2331         | 4503          | 19300     | 7000       |

The geometric parameters used in the simulations were aligned with the actual characteristics of the samples studied experimentally. Bloch–Floquet periodic boundary conditions (PBC) were applied along the *x*- and *y*-axes of the unit cell[30]. These PBCs were implemented on both faces of the unit cell to ensure consistent values of the elastic tensor components and density throughout the entire modelled structure. In the simulations, the height of the unit cell was linked to the wavelength of the acoustic wave travelling through the sample, specifically set at 20 times the wavelength of the SAW. The bottom surface of the unit cell was fixed to simulate the elliptical decay of the surface wave along the height of the sample. To determine the localization of surface modes, the intensity of individual modes was utilized as a parameter, following the relationship.

To calculate of the SAW intensity $I(f_i, q_i)$ measured in the experiment, we use the integral of the *z*-component of the displacement vector $u_z^{f_i, q_i}$ for the selected mode *i*, at the selected frequency $f_i$ and wavevector $q_i$ investigated over the free surface *A* of the studied sample[32]:

$$I(f_i, q_i) = \left| \int_A u_z^{f_i, q_i} dA \right|^2 \quad (3)$$

where $u_z$ represents the *z*-component of the total displacement *u*, defined $u = \sqrt{u_x^2 + u_y^2 + u_z^2}$ and the integration is performed over the free surface of the system.

## 4. Results

The elastic properties of different materials can be studied in various ways, including Brillouin spectroscopy. Initially, the focus was on investigating the frequency at which SAWs propagate in systems with varying CoFeB layer thicknesses. Representative Brillouin spectra of the studied multilayer systems (Fig. 1a) are shown below. The spectra clearly exhibit both Rayleigh and Sezawa SAWs (Fig. 1b-c). The dispersion relation for surface waves is presented in Fig. 1d-e. Spectra and dispersion relations for different samples with different thicknesses of CoFeB are included in Supplementary materials.

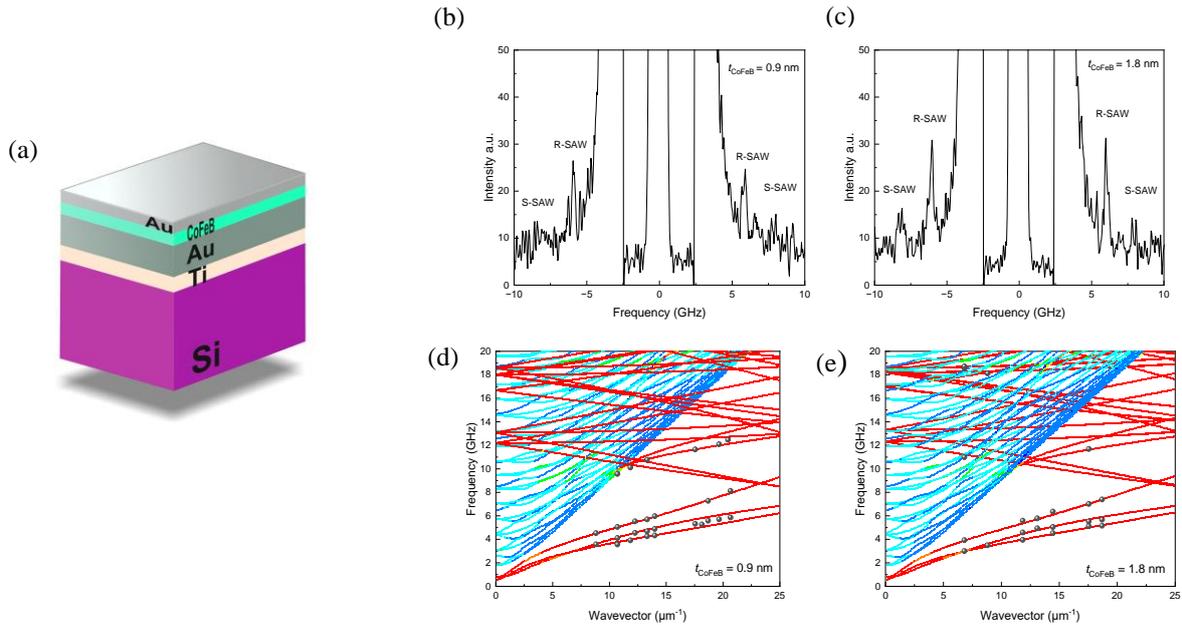

Fig. 1. Representative Brillouin spectra (b-c) of the studied materials (a), showing surface acoustic modes, including Rayleigh (R-SAW) and Sezawa (S-SAW) waves for $q = 20.44$ μm$^{-1}$ for two different CoFeB layer thickness. The dispersion relation of the investigated materials for CoFeB thicknesses $t_{CoFeB}$= 0.9 nm and 1.8 nm is shown. The frequencies of phonons extracted from experimental studies (black solid dots) and the highest intensity modes obtained from simulations (color maps) are shown as a function of the wave vector (d-e).

## 5. Discussion

The multilayer systems studied are opaque structures with significant variations in elastic parameters and density across the individual layers. The dispersion curves of surface waves are strongly influenced by the nature of the substrate, the composition of the multilayer system, and the specific properties of each layer. In general, the relative velocities of bulk transverse waves allow for the classification of layered systems into two categories: slow-on-fast and fast-on-slow[1,4,8,33,34]. In such systems, surface waves propagate through both the layer and the substrate.

It is important to note that the terms "fast substrate" and "slow layer" refer to the relationship between the velocities of bulk transverse waves propagating in the layer and in the substrate. In the slow-on-fast configuration, the presence of a slower layer on a faster substrate reduces the propagation velocity of the surface wave compared to that in the uncoated substrate. Conversely, in the fast-on-slow system, the propagation velocity of SAW increases relative to that in the uncoated substrate, as the bulk transverse wave velocity in the layer exceeds that in the substrate. Thus, a key step in characterizing these systems is determining the relative velocities of bulk waves, particularly transverse waves, within the components of the multilayer structure. This allows for the identification of the system's classification in terms of acoustic wave propagation. In the examined system, the relative velocities of the slowest bulk transverse waves follow the order: $\upsilon_{Si} > \upsilon_{CoFeB} > \upsilon_{Ti} > \upsilon_{Au}$.

This clearly indicates that the propagation velocity of waves in each layer is lower than that in the silicon substrate, allowing the system to be preliminarily classified as a slow-on-fast system (Fig. 2).

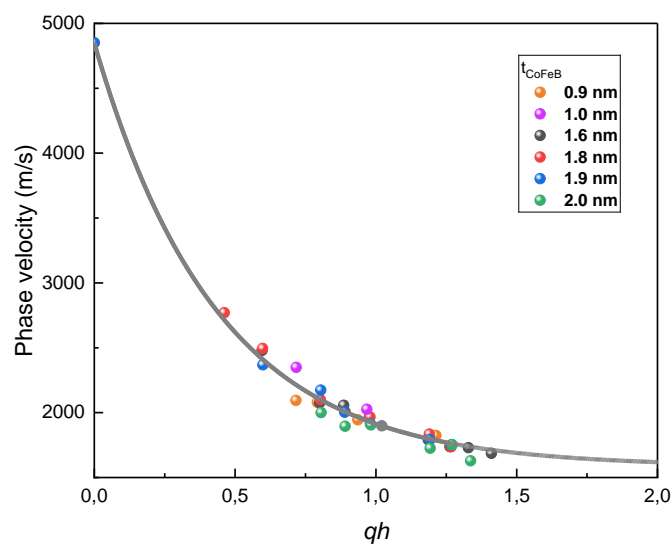

Fig. 2. Dependence of the R-SAW phase velocity on the wave number, expressed as the product of the wavevector $q$ and the total layer thickness $h$ ($h=t_{Au}+t_{CoFeB}+t_{Ti}$). Experimental points are represented by colored spheres.

The analysis of scattering by SAW depends on the characteristics of the sample. In this study, we consider a semi-infinite, homogeneous medium as well as thin layers deposited on a substrate.

Given that the penetration depth of Rayleigh waves is relatively large compared to the thickness of the layers deposited on the silicon substrate, it becomes evident that the deformation caused by SAW propagation extends throughout the entire multilayer system. This is illustrated by the wave displacement components shown in the figure below (Fig. 3).

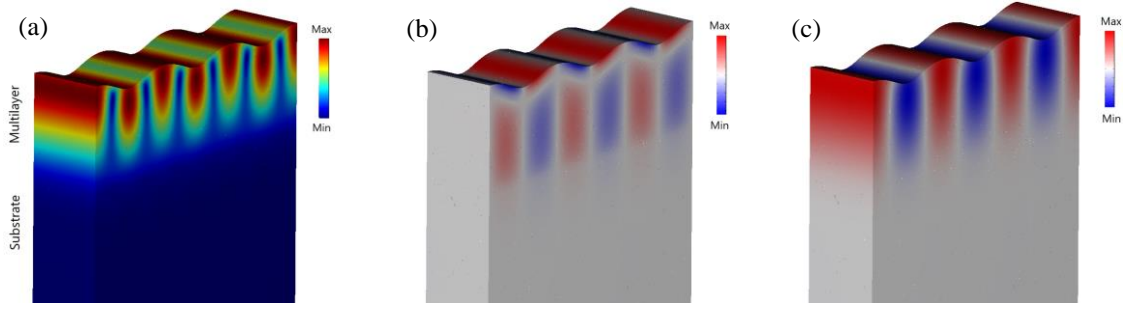

Fig. 3. The R-SAW mode profiles show the total displacement (a), the *x*-component (b) and the *z*-component (c) of the displacement for the sample with $t_{CoFeB}$ = 0.9 nm.

It can be observed that the energy associated with the propagation of the R-SAW wave is mostly concentrated within the region formed by the multilayer system. The penetration depth of the SAW in this system, estimated based on simulations, ranges from a minimum of 70 nm to a few micrometers for larger wave vectors.
Surface acoustic waves can penetrate materials to a depth of up to twice their wavelength. However, since BLS is an optical technique, the depth from which scattered light is collected is considerably shallower. Therefore, it is essential to first determine the penetration depth of light in the studied multilayer structures.

When light penetrates the sample, its amplitude decreases due to energy dissipation from absorption and scattering. This phenomenon is quantified by the extinction coefficient, which depends on the refractive index and particle size. The extinction coefficient directly influences the depth of material penetration by light waves[21]. This light penetration depth can be calculated using the following formula:

$$\delta = \frac{\lambda_0}{4\Pi\kappa} \quad (3)$$

where $\lambda_0$ is the wavelength of the laser (532 nm), and $\kappa$ is the extension coefficient characteristic for each layer. Table 2 provides information on the refractive index, extinction coefficient, and light wave penetration depth for each layer according to equation 3. The effective refractive index, extinction coefficient, and light wave penetration depth for CoFeB were determined based on the chemical composition of the sample, where cobalt, iron, and boron are present in proportions of 20%, 60%, and 20%, respectively.

Table 2: The depth of light wave penetration of the materials presents in the Si/Ti/Au/CoFeB/Au sample.

| Material | Si | Ti | Au | CoFeB | Au |
|---|---|---|---|---|---|
| Refractive index (n) | 4.1520 | 2.4793 | 0.5439 | 2.2429 | 0.5439 |
| Extinction coefficient ($\kappa$) | 0.0518 | 3.3511 | 2.2309 | 2.5683 | 2.2309 |
| Light penetration depth (nm) | 817 | 13 | 19 | 35 | 19 |

Considering the light penetration depth values in individual layers, an approximate method can be applied to determine the overall light penetration depth in the studied system. The average penetration depth for the entire system will be calculated as a weighted average of the penetration depths of each layer, considering the thicknesses of the layers. For each layer, we will multiply the penetration depth by the thickness of the layer, sum these values, and then divide by the total thickness of the system, according to the following equation:

$$\delta_{eff} = \frac{\sum(\delta_n \cdot d_n)}{\sum d_n} \tag{5}$$

where: $\delta_n$ – penetration depth of the n-th layer, $d_n$ - thickness of the n-th layer.

Thus, the penetration depth of light in the studied material is approximately 19 nm (for 0.9 nm CoFeB thickness). This directly results from the application of the contribution ratio method based on Table 2. Light will penetrate the system the least when the effective penetration depth is at its minimum, which typically occurs with materials with the highest extinction coefficient (κ) in the first layer through which light passes.

The penetration depth of SAWs in materials can reach up to two SAW wavelengths. However, due to the optical nature of the BLS technique, the actual depth from which the scattered light is collected is significantly smaller. Consequently, treating the multilayer system as an effective layer is valid within the region where light is collected or as one effective layer. The main question is which of these approaches is correct. To answer this, we examine two situations: the first treats multilayers as a single effective layer with effective elastic parameters, while the second approach defines the elastic parameters only in the region where light penetrates. For the calculation of the dispersion relation in both approaches, we use the parameters displayed in Fig. 4.

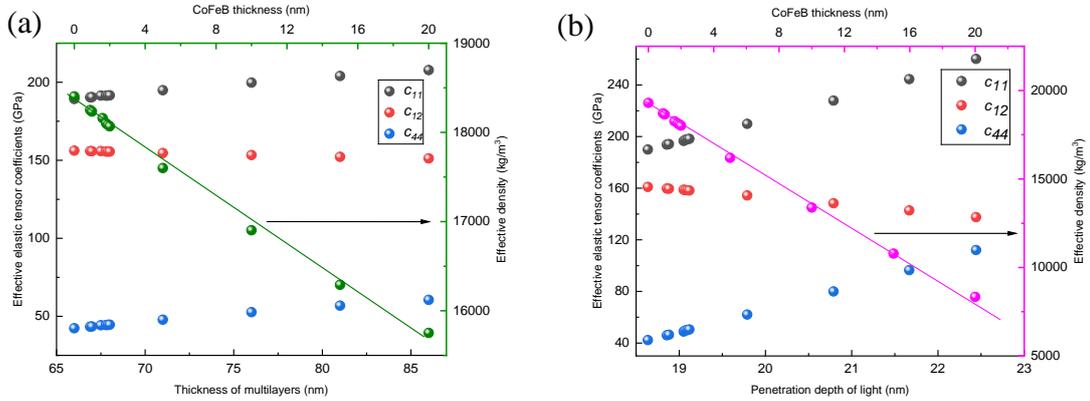

Fig. 4. Effective elastic parameters and density for the multilayers are treated as a single effective layer (a) and for the case where the multilayer is considered effective in the region of light penetration (b).

The effective elastic tensor coefficients $c_{ij}^{eff}$ for a multilayer system were calculated using the following equation:

$$c_{ij}^{eff} = \frac{1}{h}\sum_n h_n c_{ij}^{(n)} \tag{6}$$

where: $c_{ij}^{eff}$ represents the effective coefficient of the elastic tensor, $h_n$ is the thickness of the n-th layer, $c_{ij}^{(n)}$ is the elastic tensor coefficient specific to the n-th layer, $h$ is the total thickness of the multilayer system, defined as the sum of all individual layer thicknesses:

$$h = \sum_n h_n \quad (7)$$

Formula 6 essentially calculates a thickness-weighted average of the elastic coefficients across all layers in the system or selected layers. The $c_{ij}$ values calculated to satisfy the Born stability criteria for elasticity tensors[35]. These are fundamental conditions that must be met for a material to be mechanically stable, ensuring that the strain energy of the material is positively defined. For cubic symmetry, which is indirectly influenced by the substrate symmetry, these conditions are expressed as follows: *$c_{11}$ > 0, $c_{44}$ > 0, $c_{11}$ - $c_{12}$ > 0, $c_{11}$ +2 $c_{12}$ > 0*. The density of the effective layer was calculated analogously.

The calculated effective elastic tensor and density with respect to the thickness of the CoFeB layer are shown in Fig. 4. To generalize our observation, we calculated the elastic tensors and densities for CoFeB thicknesses ranging from 0 nm to 20 nm. The change in CoFeB thickness affects the calculated parameters when treating the multilayer system as a single effective layer. Depending on the penetration depth of the light, the values of the elastic tensor and density also vary. It illustrates that the values of $c_{ij}$ and density are more sensitive to changes in the CoFeB thickness when only considering the thickness of the region penetrated by light.

By utilizing the elastic parameters and density presented in Table 1 and Fig. 4, we calculate the phase velocity of the SAW for a sample with 0.9 nm of CoFeB, as shown in Fig. 5.

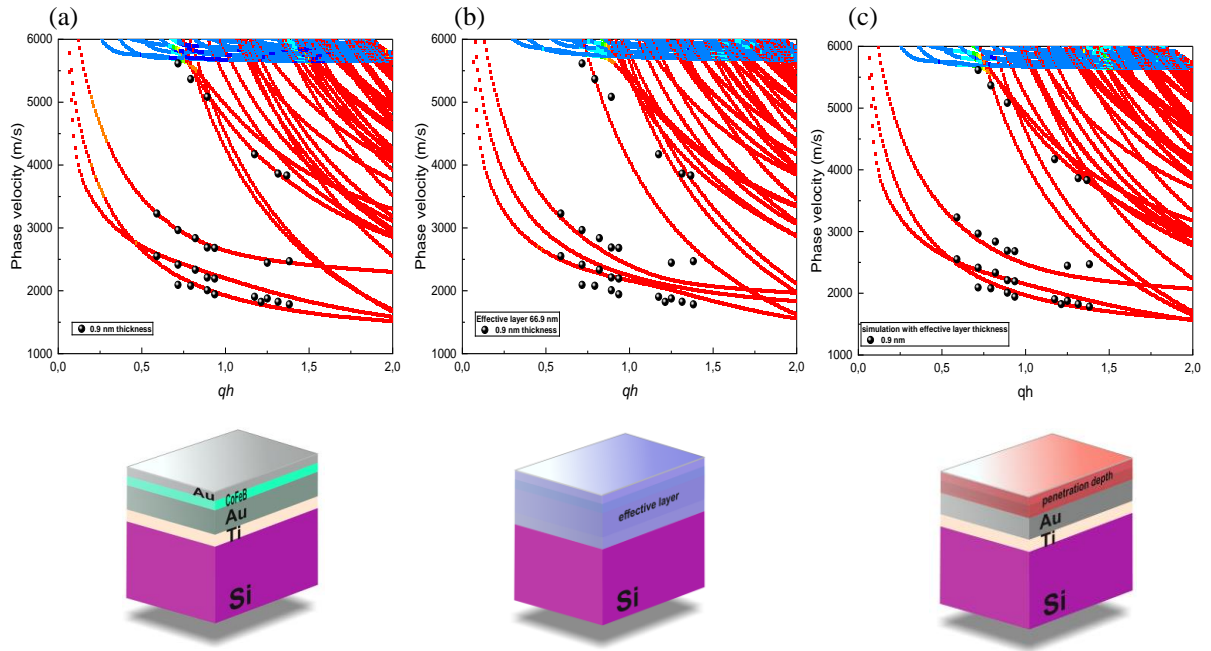

Fig. 5. Phase velocity dispersion of SAWs measured for the S/Ti/Au/CoFeB ($t_{CoFeB}$ = 0.9 nm)/Au sample, plotted as a function of wave number, considering three models: (a) the sample treated as a multilayer, (b) the sample treated as a single effective layer (blue region in the schematic), and (c) the sample treated as an effective layer limited to the light penetration depth (19 nm, red region in the schematic). Black points represent experimental data, while colored points are derived from FEM simulations using the parameters given in Table 1 and Fig. 4. Schematic representations of the three modeling approaches are shown below the corresponding phase velocity graphs.

The agreement between simulation and experimental results is better when the multilayer system is treated as an effective layer within the light penetration region rather than as a single effective layer (Fig. 5b and 5c). This approach demonstrates that the capping layers have a significant impact on SAW propagation velocity. However, it is essential to consider the light penetration depth within the given multilayer system (Fig. 5c). A comparison of different approaches indicates that, even when treating the system as a multilayer, it can still be regarded as an effective layer within the light penetration region. This becomes evident when comparing the agreement between simulations and experimental results in the figure above. Notably, the poorest fit occurs when the entire multilayer system is treated as a single effective layer (Fig. 5b).

For the studied systems, Young's modulus was determined in two different scenarios (Fig. 6). All calculations were performed using the equations 8-11 in the first approach for the multilayer system treated as a single effective layer. In the second approach, the effective layer was considered only within the laser light penetration region—approximately 19 nm for the sample with a 0.9 nm CoFeB layer thickness. Young's modulus, $E$, can be calculated using the following equation[36,37]:

$$E = \frac{1}{s'_{33}} \quad (8)$$

where $s'_{33}$ is the stiffness component in the system's measurement plane (expressed in Voigt notation). For a material with cubic symmetry, this component is defined as follows[38]:

$$s'_{33} = s_{11} - \left(s_{11} - s_{12} - \frac{1}{2}s_{44}\right)[1 - cos^4\varphi sin^4\psi - sin^4\varphi sin^4\psi - cos^4\psi] \quad (9)$$

where $s_{11}$, $s_{12}$, $s_{44}$ represent the stiffness components in the system's crystal plane, while $\varphi$ and $\psi$ are arbitrary rotation angles used to transform the cubic crystal coordinate system into a lattice plane system. The relationships between the stiffness components $s_{11}$, $s_{12}$, $s_{44}$ and the elastic constants $c_{11}$, $c_{12}$, $c_{44}$ are given as follows:

$$s_{11} = \frac{c_{11} + c_{12}}{(c_{11} - c_{12})(c_{11} + 2c_{12})}$$

$$s_{12} = \frac{-c_{12}}{(c_{11} - c_{12})(c_{11} + 2c_{12})} \quad (10)$$

$$s_{44} = \frac{1}{c_{44}}$$

By applying Eqs. (8)–(10), Young's modulus $E$ can be expressed as:

$$E = \frac{2(c_{11}-c_{12})(c_{11}+2c_{12})c_{44}}{2(c_{11}+c_{12})-(c_{11}+2c_{12})(2c_{44}-c_{11}+c_{12})[1-\cos^4\varphi\sin^4\psi-\sin^4\varphi\sin^4\psi-\cos^4\psi]} \quad (11)$$

This equation is applicable to a cubic crystal system and considers the variation of Young's modulus with respect to the wave propagation direction within the crystal. Additionally, we utilized ELATE[39], an open-source online tool for elastic tensor analysis, to derive Young's modulus from the given elastic tensor.

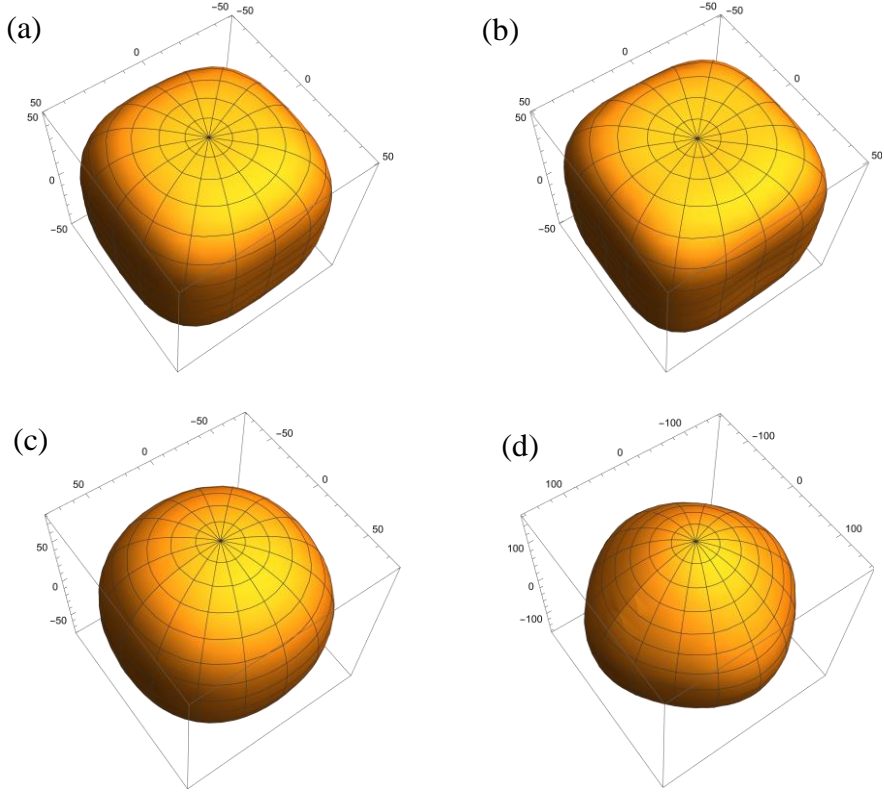

Fig. 6. The 3D view of the calculated Young's modulus ($E$) for different CoFeB thicknesses in the sample: $t_{CoFeB} = 0.9$ nm (a, b) and $t_{CoFeB} = 20$ nm (c, d) treated as an effective layer (a, c) and only within the penetration depth region (b, d).

As seen in Fig. 6, the anisotropy of the Young's modulus changes with increasing CoFeB thickness. This effect is observed both when treating the multilayer system as a single effective system and when calculating the Young's modulus within the light penetration region. The variation in Young's modulus with respect to the crystallographic direction can be determined using the elastic anisotropy of the crystalline material, known as Zener anisotropy (A). Zener anisotropy for cubic structure can be calculated by the equation[38,40–43]:

$$A = \frac{2c_{44}}{c_{11}-c_{12}} \quad (12)$$

where $c_{11}$, $c_{12}$ and $c_{44}$ are the elastic constants. The results obtained for the elastic parameters and Young's modulus are presented in Table 3.

Table 3. The Zener anisotropy values (A) for the samples with different CoFeB thicknesses, considering both for effective layer and light penetration depth.

| CoFeB thickness (nm) | Thickness of effective layer (nm) | A | Penetration depth of light (nm) | A |
|---|---|---|---|---|
| 0 | 66 | 2.56 | 18.64 | 2.92 |
| 0.9 | 66.9 | 2.52 | 18.86 | 2.71 |
| 1 | 67 | 2.51 | 18.88 | 2.69 |
| 2 | 68 | 2.47 | 19.12 | 2.52 |
| 5 | 71 | 2.38 | 19.79 | 2.24 |
| 10 | 76 | 2.27 | 20.79 | 2.01 |
| 15 | 81 | 2.19 | 21.67 | 1.90 |
| 20 | 86 | 2.14 | 22.44 | 1.83 |

From an application perspective, these studies are significant as they enable the control of surface wave frequency by adjusting the thickness of the magnetic component in the layered material. This allows for precise tuning of the elastic and dynamic properties of multilayer structures, which is crucial for modern acousto-electronic and spintronic technologies. The ability to manipulate the interaction between acoustic and spin waves through structural modifications has significant implications for next-generation communication and computing technologies. For example, controlling SAW frequency enables optimization of resonators and surface wave filters, which are fundamental components in RF (radio frequency) and microwave devices used in wireless communication, radar systems, and high-frequency signal processing[44]. Additionally, the ability to adjust the thickness of effective layers in a multilayer structure affects the penetration depth of SAWs, determining how much of the wave energy is confined to the surface or distributed deeper into the material. This aspect is critical for enhancing wave confinement, minimizing energy loss, and improving device sensitivity. Ultimately, these studies not only contribute to the fundamental understanding of SAW behavior in complex layered materials but also drive innovations in advanced acousto-electronic and spintronic devices, paving the way for more efficient and compact technologies in telecommunications, sensing, and computing.

## 6. Conclusions

The study reveals the critical influence of CoFeB layer thickness on surface acoustic wave (SAW) propagation characteristics. By examining variations in layer thickness from 0.9 nm to 2 nm, we observed significant changes in the propagation of Rayleigh (R-SAW) and Sezawa (S-SAW) waves, highlighting the sensitivity of acoustic wave behavior to nanoscale structural modifications. Light penetration depth analysis demonstrated that accurate modelling requires consideration of only an approximately 19 nm thick layer. This insight significantly improves the precision of effective medium modelling, addressing a crucial limitation in previous experimental approaches. Neglecting light penetration depth can lead to inaccurate estimations of acoustic wave velocities, emphasizing the importance of considering this parameter for obtaining reliable and reproducible results.

The Zener anisotropy coefficient of the studied samples exhibits a systematic decrease with increasing CoFeB layer thickness. This trend indicates a progressive homogenization of the material's elastic properties and confirms the moderate elastic anisotropy of CoFeB, providing deeper insights into the material's structural evolution at the nanoscale.

The proposed methodology, which integrates finite element method numerical simulations with Brillouin light scattering experimental measurements, offers a robust and precise toolkit for predicting acoustic wave behavior in complex multilayered structures. This approach represents a significant advancement in characterizing advanced material systems. This phenomenon opens novel opportunities for manipulating acoustic wave properties in spintronic and acoustic devices, where precise control of wave propagation is paramount.

**Acknowledgement:**

This work was supported by the Polish National Science Centre under grant no: UMO - 2020/37/B/ST3/03936